\documentclass[preprintnumbers,amsmath,amssymb,twocolumn]{revtex4}
\usepackage{graphicx}
\usepackage{dcolumn}
\usepackage{bm}

\begin{document}

\title{Diamonds levitating in a Paul trap under vacuum : \\ measurements of laser-induced heating via NV center thermometry}

\author{T. Delord$^{1}$}
\author{L. Nicolas$^{1}$}
\author{M. Bodini $^{1}$}
\author{G. H\'etet$^{1}$} 

\affiliation{$^1$Laboratoire Pierre Aigrain, Ecole normale sup\'erieure, PSL Research University, CNRS, Universit\'e Pierre et Marie Curie, Sorbonne Universit\'es, Universit\'e Paris Diderot, Sorbonne Paris-Cit\'e, 24 rue Lhomond, 75231 Paris Cedex 05, France.}

\begin{abstract}
We present measurements of the Electronic Spin Resonance (ESR) of Nitrogen Vacancy (NV) centers in diamonds that are levitating in a ring Paul trap under vacuum. We observe ESR spectra of NV centers embedded in micron-sized diamonds at vacuum pressures of $2\times10^{-1}$ mbar and the NV photoluminescence down to $10^{-2}$ mbars. Further, we use the ESR to measure the temperature of the levitating diamonds and show that the green laser induces heating of the diamond at these pressures.
We finally discuss the steps required to control the NV spin under ultra-high vacuum.
\end{abstract}

\maketitle

Engineering the motional state of massive oscillators will be an important step forward for modern quantum science \cite{Aspelmeyer}. 
Hybrid-opto-mechanical schemes, where the center of mass of oscillators is coupled to single atoms \cite{Rabl}, have been propounded to harness this challenge and
levitating nano-objects proposed as a viable experimental platform \cite{Chang19012010, Yin}. 
It is indeed possible to benefit from the inherent decoupling of levitating particles internal degrees of freedom from the surrounding environnement and to cool its center of mass mode close to the motional ground state \cite{Chang19012010, Yin}. Using hybrid-opto-mechanical schemes with single atoms coupled to or embedded in levitating particles would not only also enable ground state cooling, but also preparing macroscopic non-gaussian motional states and Schr\"{o}dinger cat states \cite{Yin} or to perform matter wave interferometry \cite{Scala, Romero, Yin2}.
To this end, it is important to operate under low vacuum to minimize collisions with air particles, which prevent the center of mass from reaching low temperatures.  
Thus far, experiments using NV centers embedded in optically levitating diamonds show heating induced by the trapping laser \cite{Rahman, Neukirch, Hoang}, so that high purity diamonds \cite{Frangeskou} or higher trapping laser wavelengths \cite{Hoang} need to be used to mitigate this effect. 
In optical traps, electronic spin read-out of NV centers was observed at tens of millibars of pressure \cite{Hoang}, beyond which diamonds are lost from the trap.

A solution to particle heating is to use scattering-free traps such as Paul traps \cite{Kuhlicke, Nagornykh, Alda, Millen} or magneto-gravitational traps \cite{Hsu}. 
In \cite{Delord, Neukirch}, the photoluminescence of NV centers in a Paul trap was observed under ambient conditions and the angle stability of the particle was demonstrated using Electronic Spin Resonance (ESR) in \cite{Delord}.
In this paper, we report measurements of the electronic spin resonance of NV centers embedded in diamonds that are levitating in an ion trap under vacuum.  Using a slightly modified set-up compared to \cite{Delord}, featuring a small copper ring employed both for trapping and microwave excitation, we could observe the photoluminescence of NV centers at pressures close to $10^{-2}$ mbars for more than 40 minutes. We also detect ESR signals down to $2\times 10^{-1}$ mbars and use it to infer the temperature of the levitating diamond.

\begin{figure}[ht!]
\centerline{\scalebox{0.19}{\includegraphics{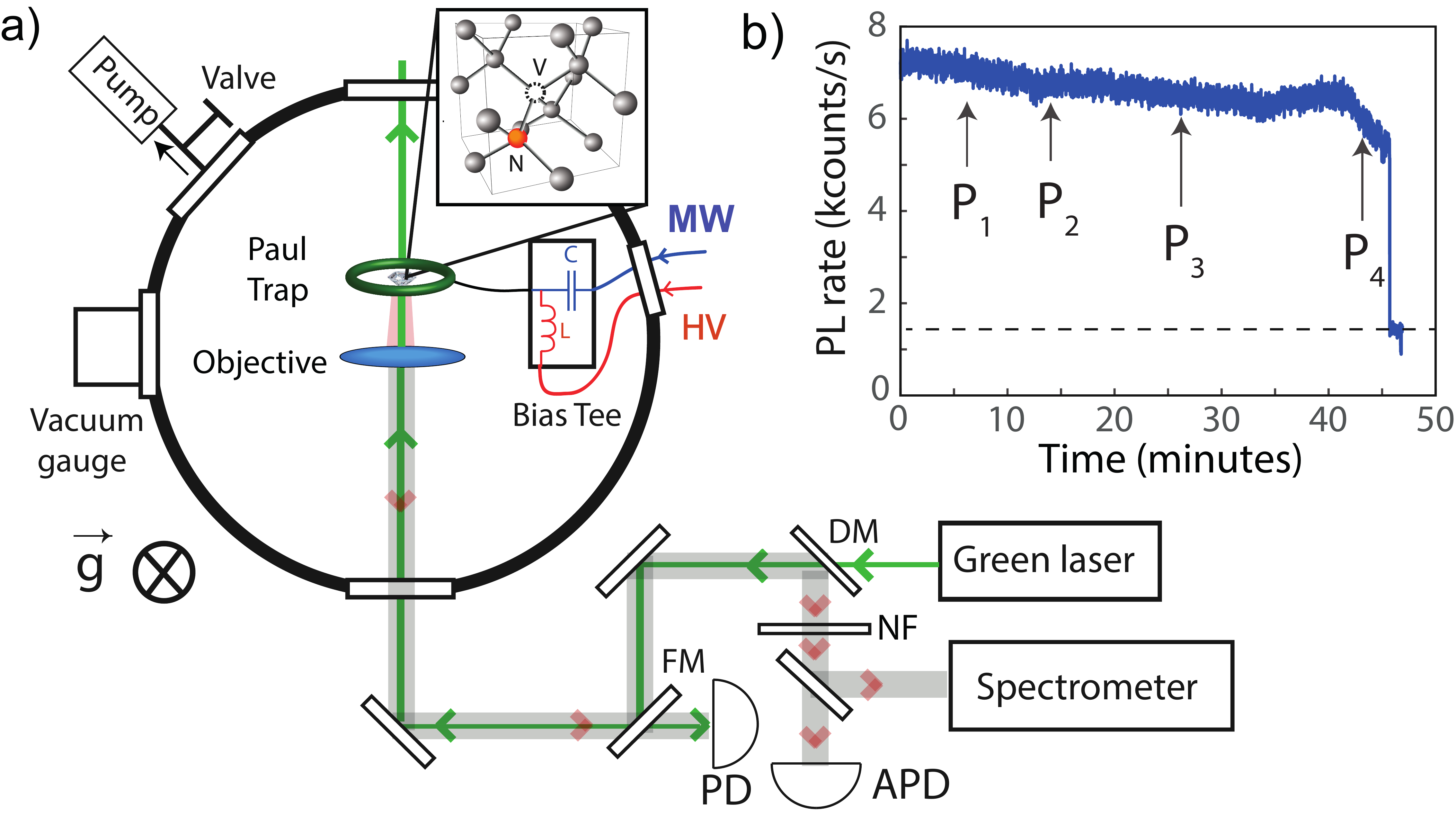}}}
\caption{a) Schematics of the experimental apparatus. The Paul trap, objective, and Bias Tee are enclosed in a vacuum chamber. 
A green laser is focused onto a diamond levitating in the Paul trap.  The photoluminescence from the NV centers in the trapped diamond is collected by the same objective and measured either on an avalanche photodiode (APD) or on a spectrometer. 
DM= dichroic mirror, NF=Notch filter centered at 532 nm. FM=flipping mirror. PD=Photodetector, HV=High Voltage, MW= Microwave. b) Photoluminescence (PL) level as a function of time as the pressure in the chamber decreases. Intermediate measurements of the pressure levels are indicated. $P_{1,2,3,4}=(1.6, 1.4 ,1.2, 1.0) \times 10^{-2}$ mbars.  The dashed line denotes the background level. 
}\label{Setup}
\end{figure}

The experimental set-up is depicted in Fig.~1- a). The trap and the objective are enclosed in a vacuum chamber. The trap is mounted on a 3-axes translation stage.
Compared to \cite{Delord}, the employed ion trap is a ring trap in a Paul-Straubel configuration \cite{Straubel,Yu}. 
It consists in a 300 $\mu$m thick copper wire with an inner radius of 700 $\mu$m.
With micron sized diamonds, we can operate the trap with a peak-to-peak voltage ranging from $V_{\rm ac}$=100~V to 4000~V at driving frequencies in the kHz range. Choosing a ring geometry rather than the needle trap used in \cite{Delord} implies that the trap itself can be used to generate the oscillating magnetic field.
This avoids extra control of the antenna position and asymmetries in the trapping potential produced by the remote wire antenna.
The micro-wave excitation was done via a Bias Tee, as depicted in Fig.~1-a). Microwave powers of 0.5~W are used to excite the NV centers' electronic spins.
Further details about the diamonds used, the number of NV centers, the loading protocol and the imaging system can be found in \cite{Delord}.

Here, the voltage needs to be lowered to 600~V to avoid plasmas that would otherwise appear at around 10 mbars in the chamber. 
This means that careful preselection of particles with a high charge to mass ratio need to be performed at atmospheric pressures beforehand. To do so, we operate the trap at 4000~V, inject diamonds with a 10 $\mu$m diameter, and systematically measure the onset of instability. 
In Paul traps, stability is fulfilled if the trapping frequency is larger than than the secular frequency \cite{Pau90}. 
Measuring the trapping frequency at which instability takes places provides a means to estimate the charge to mass ratio. 
When this frequency is below 1 kHz at 4000~V, we estimated that the trapping time was too short at 600~V and at low pressures so another particle is loaded. 
This mandatory preselection highlights another benefit of the ring trap where the trap anharmonicities \cite{Yu2} do not eject the particle close to the instability threshold. The whole procedure typically requires 3 to 4 loading steps before a high enough charge to mass ratio is attained. 

Once a particle is trapped, the voltage and frequency are lowered in air, following an iso-$q$ curve, $q$ being the stability parameter of the trap at 1 bar \cite{Pau90}. Once 600~V is reached, the turbomolecular pump is turned on.
As observed both in transmission and by looking at the back-scattered green light, when the pressure reaches 500 mbars already, the center of the trap shifts. The back-reflected scattered beam size thus reduces significantly (by about a factor of 5 at 10 mbars of vacuum pressure) due to a displacement of the particle with respect to the objective focal point. This takes place because the secular frequency depends on the damping rate, {\it i.e.} on the vacuum level \cite{Izmailov, Hasegawa1995}. In the presence of residual electric fields or gravity, changing the confinement via the damping rate displaces the particle.
The diamond back scattered image also appeared elongated in a direction perpendicular to the optical axis, consistant with a large micromotion amplitude. When the damping rate decreases the particle indeed explores a larger volume away from the center of the trap  so that micromotion increases. 
When the vacuum reaches 10$^{-2}$ mbars, the voltage can be increased back to 4000 V without any arching in the chamber. In principle, the optimum voltage for high confinement and no arching, could be chosen by following the Paschen law \cite{Paschen}.  For the present measurements however, we kept the particle at a voltage of 600~V because changing the trapping frequency below 1 mbars often imparts kicks to the diamond that are large enough to make it leave the trap. 

The photoluminescence (PL) of the NV centers is collected using the confocal microscope described in Figure~\ref{Setup}-a).
The PL signal is then filtered using a Notch filter centered at 532 nm and can be directed either onto an avalanche photodiode or a spectrometer.
With a mW of laser excitation, we can collect around $10^5$ counts per second on the avalanche photodiode at atmospheric pressure.  As the pressure decreases however, the PL decreases due to the aforementioned damping-induced micromotion and position shift, and possibly to the high diamond temperature at low pressures as we will see.
Fig.~1-b) shows a time trace of the photoluminescence taken at pressure levels starting from $1.8\times10^{-2}$ and reaching 10$^{-2}$ mbars.
The PL rate is about 8 kcounts/s at $2\times 10^{-2}$ mbars and here drops at 47 minutes.
The reason for the observed change in the PL over long time scales can be attributed to instabilities in the trap frequency, which displaces the particle from the focal spot.
A more rapid drop of the PL is seen at 42 minutes, when the pressure reaches 10$^{-2}$ mbars, and then the particle leaves the trap abruptly at 47 minutes, as confirmed by monitoring the retro-reflected beam.
This experiment was repeated more than 20 times, and after reaching 10$^{-2}$ millibars, all particles leave the trap after a couple of minutes. Before being lost, they shake in a random fashion, the rate of which was found to depend upon the laser power. This sudden escape of the diamond from the trap therefore seems to be induced by the green laser.
We will now show with the ESR that the diamond temperature increases significantly at pressure levels below the mbar due to the green laser using NV center thermometry.

\begin{figure}[ht!]
\centerline{\scalebox{0.21}{\includegraphics{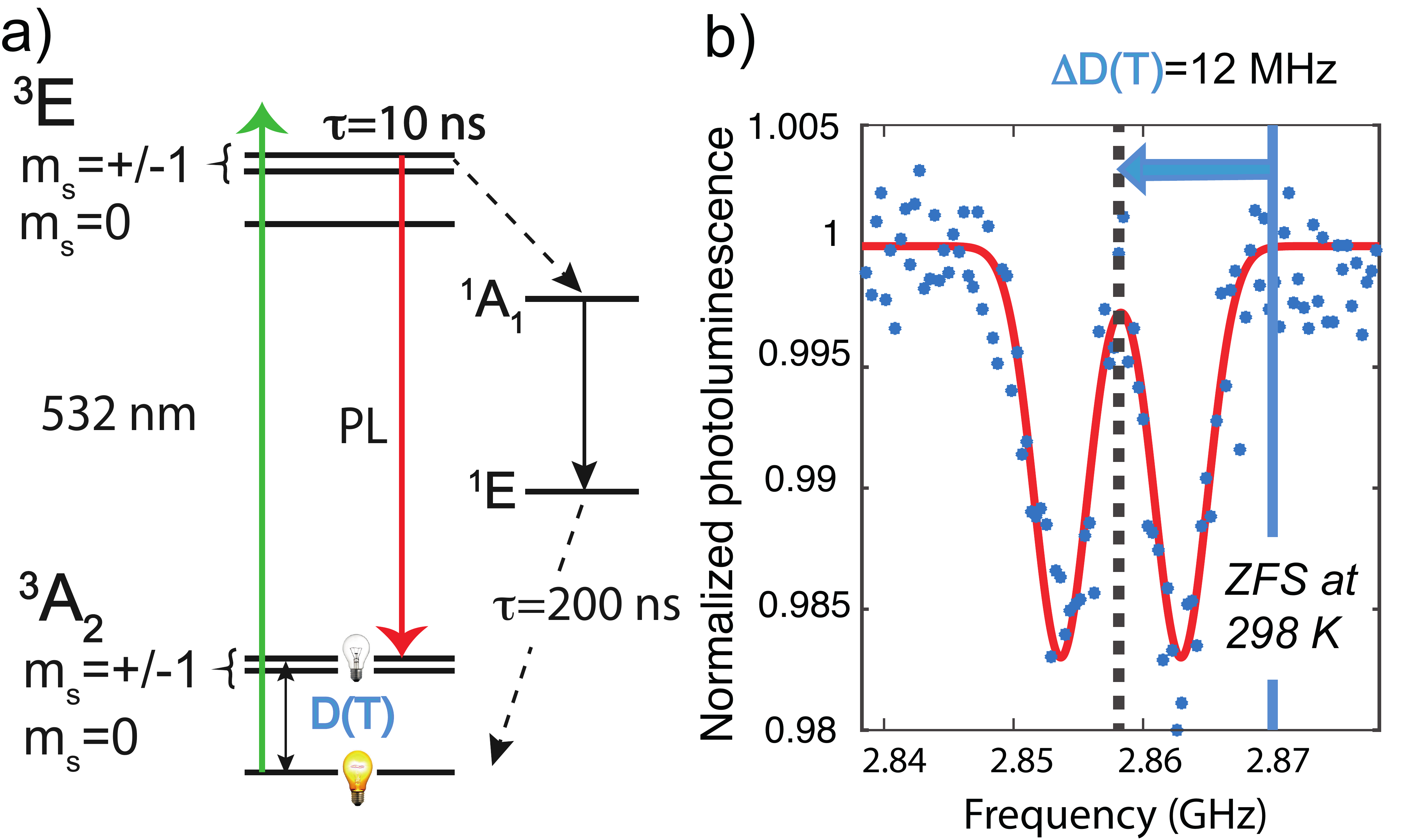}}}
\caption{a) Level scheme of the NV$^-$ center. PL=Photoluminescence. The two on/off light bulbs denote the bright/dark electronic states. $D(T)$ is the temperature dependent zero-field splitting (ZFS).  $D$ = 2.87 GHz at 298 K. b) Electronic spin resonance spectrum taken at 1 mbar of vacuum pressure and 450 $\mu$W of green laser power. A frequency shift $\Delta D(T)=12$ MHz from the ZFS at ambient conditions is observed. The solid line is a double-Gaussian fit.
}\label{Spectrum}
\end{figure}

We probe the NV$^-$ centers' electronic spin transitions in the trapped diamond using a microwave tone.
The level structure of the NV$^-$ spin is depicted in Figure~\ref{Spectrum}-a).
The NV$^-$ has two unpaired electrons so the ground state is a spin triplet. The degeneracy between the $|m_s=0\rangle$ and $|m_s=\pm 1\rangle$ manifolds, defined with respect to the NV center axis, is lifted by $D=2.87$ GHz due to spin-spin interaction, a value that is highly temperature dependent \cite{Acosta}.
Assuming a similar environnement for all NV centers, the Hamiltonian of the NV center in the $^3A_2$ ground state manifold reads
\begin{eqnarray}
\mathcal{H}=D S_z^2+E(S_x^2-S_y^2)+\sum_i \vec{\mu_i}\cdot \vec{B},
\end{eqnarray}
where $D$ is the energy splitting between the $|m_s=0\rangle$ and $|m_s=\pm 1\rangle$ states in the absence of magnetic field and $E$ is the splitting between the $|m_s=\pm 1\rangle$
due to the broken axial symmetry of the NV centre. 
$\sum_i \vec{\mu_i}\cdot \vec{B}$ is the coupling between the spins $i$ with magnetic moment $\vec{\mu_i}$ with the magnetic field $\vec{B}$.

As depicted in Fig 2. a), due to an intersystem-crossing in the excited state, scanning the frequency of a microwave tone around 2.87 GHz under green excitation results in a drop of the photoluminescence from $S_{max}$ to $S_{min}$ \cite{Gruber}.
For small magnetic fields, this Hamiltonian results in a double peak resonance spectrum, $D$ being the centre of the double peaks, which is about 2.87 GHz at room temperature. 

\begin{figure}[ht!]
\centerline{\scalebox{0.35}{\includegraphics{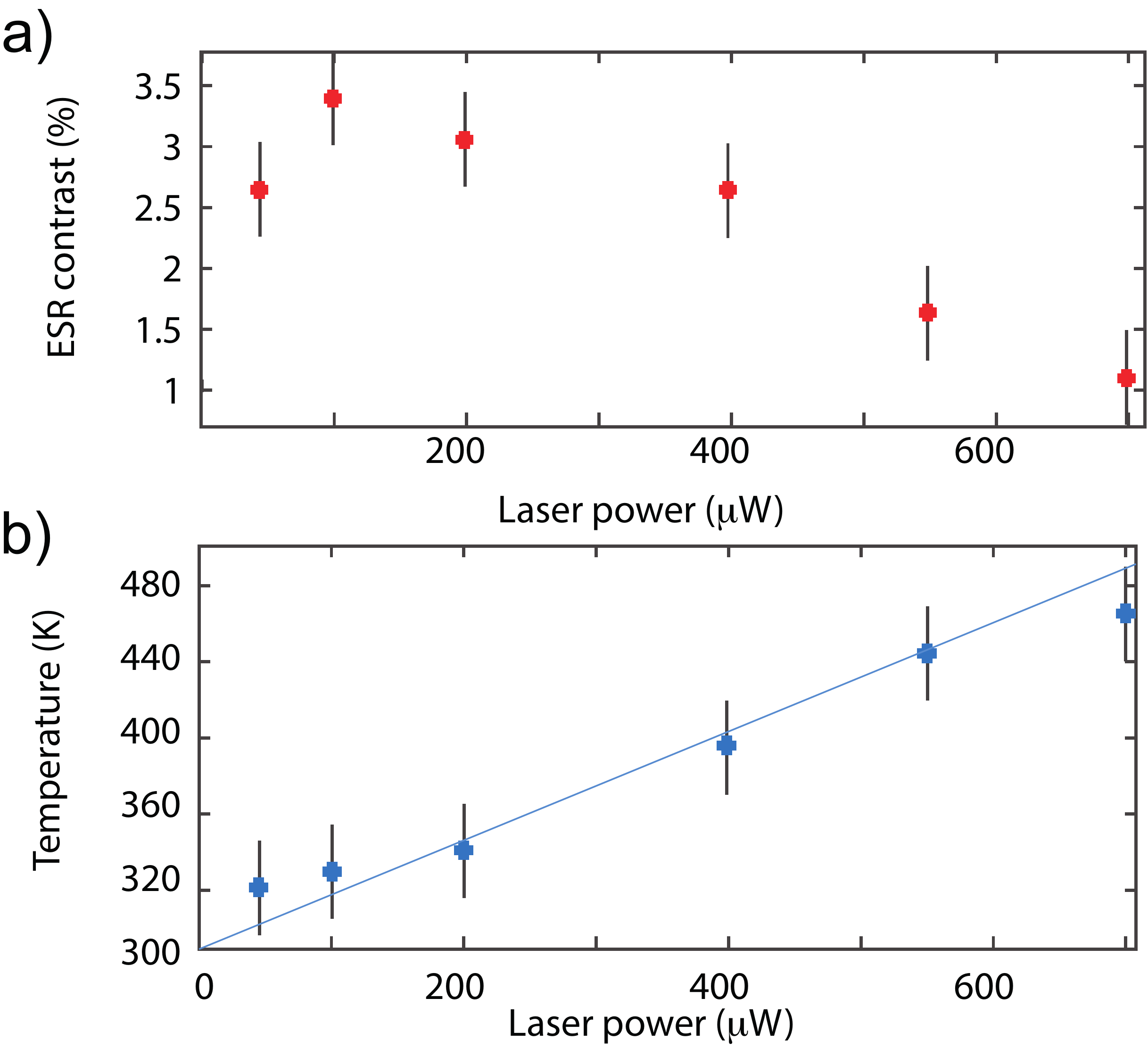}}}
\caption{Electron spin resonance contrast of NV$^-$ center (a) and temperature of a levitating diamond (b) as a function of laser power at 1 mbar of vacuum pressure. The solid line is a linear fit with origin at 298 K.
}\label{Power}
\end{figure}

To manipulate the electronic spin, we use the ring trap itself as the antenna. A Bias Tee (see Fig.~1.a)) is used to mix the low frequency high voltage signal used for trapping and the signal from the microwave generator. 
Under ambient conditions, the ESR contrast, taken to be the normalized depth of the ESR dip $(S_{max}-S_{min})/(S_{max})$, was measured to be 3.5\%.
Figure~\ref{Spectrum}-b) shows a typical ESR spectrum taken at 1 mbar of vacuum pressure and 450 $\mu$W of green laser power. 
For this data, an ESR contrast of 1.7\% is observed with a zero-field-splitting (ZFS) of 2.8558 GHz, corresponding to a shift of 12 MHz from the room temperature splitting.  

Fig. \ref{Power}-a) shows measurements of the ESR contrast for the same particle as a function of laser power, at 1 mbar of vacuum pressure. The ESR contrast starts at 2.5 \% at 50$\mu$W and reduces to 1\% at 700 $\mu W$. 
Let us note that, in the range of laser powers used in the experiment, no change in the ESR contrast or width was observed under atmospheric conditions \cite{Jensen, Singam}.
The drop in the contrast can thus be attributed to an increased temperature of the diamond.
Indeed, as the pressure is reduced, there is no cooling due to exchange of heat with the residual gaz molecules any longer. The temperature of the diamond can thus rise significantly. This happens when the mean free path of gas molecules is larger than the size of the diamond, {\it i.e} for a Knudsen number above unity. 
Here, the decrease in the ESR contrast with temperature is due to nonradiative processes that quench the NV center's
fluorescence spin readout. It was indeed shown that multi-phonon processes reduce the lifetime of the optically excited $|m_s=0\rangle$ state, thereby limiting the ESR contrast \cite{Toyli}.  Our results differ from the results obtained in \cite{Hoang} where an increase in the contrast as a function of pressure was observed in an optical trap. It was argued that heating removes low-quality NV$^-$ centres near the surface \cite{Hoang}. Here, the diamonds are two orders of magnitude larger, which may explain the opposite dependence of the contrast with temperature. 

\begin{figure}[ht!]
\centerline{\scalebox{0.57}{\includegraphics{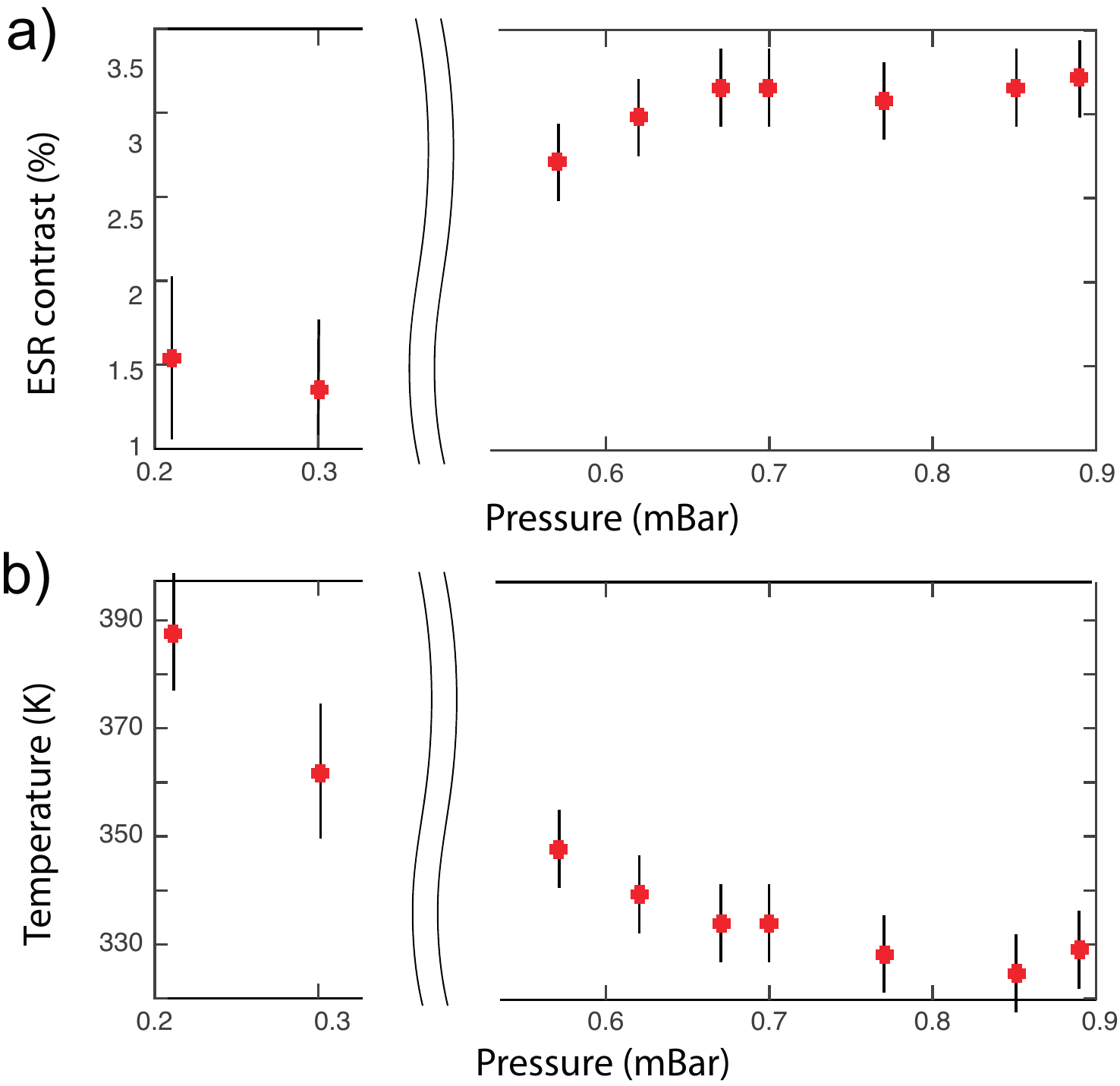}}}
\caption{Electron spin resonance contrast of NV$^-$ center (a) and temperature of a levitating diamond (b) as a function of vacuum pressure at 40$\mu$W of green laser pump power.}\label{Pressure}
\end{figure}

Accompanying this change in the contrast, a pronounced shift of the ZFS is observed in the ESR shown in Fig.~2-b).
Since the lattice extension changes with temperature, the factor $D$ depends on temperature. This shift in the ZFS therefore confirms the interpretation that the ESR contrast drops due to temperature \cite{Acosta, Hoang, Toyli}. 
$D$ was accurately described by a third-order polynomial between 300 K and 700 K in \cite{Toyli}
\begin{eqnarray}
D(T)=a_0+a_1 T^1 + a_2 T^2 +a_3 T^3,
\end{eqnarray}
where $a_0=2.8697$ GHz,
$a_1=9.7\times 10^{-5}$ GHz/K, and $a_2=-3.7\times 10^{-7}$GHz/K$^2$
which yields shifts of about 80 kHz/K close to 300 K.
Using these values, we can deduce the temperature of the levitating diamond from the frequency shift of the ZFS.
Figure \ref{Power}-b) shows the dependence of temperature on laser power. 
A pronounced monotonic increase in the temperature with laser power is observed, reaching up to 470K at 700 $\mu$W of green laser power. This heating effect may also explain the sudden loss of particle at lower vacuum pressures.

Absorption of light in the material and subsequent heating can be attributed to nitrogen defects inside the diamond since nitrogen is naturally present in commercially available HPHT diamond powders \cite{Frangeskou}.  
The solid line shows a linear fit to the data, the slope of which depends on the absorption of the diamond at 532 nm, on the particle size, and on the micro-motion amplitude. Note that because of micromotion these laser power values cannot be simply related to the actual intensity at the diamond location and can only be taken as indicative numbers.

We now turn to measurements realized at lower vacuum pressures and with 40 $\mu$W of laser power. Figure \ref{Pressure} shows the dependence of ESR contrast and temperature with pressure in the chamber in the 0.2 to 0.9 mbars range. 
These measurements were done with the same particle in the trap as for the measurements shown in Figure \ref{Power}. 
Due to the faster loss rate of the particle when the pressure lies in the $10^{-1}$ mbars range, less data points were taken when approaching these pressures.
The ESR contrast reduces from 3.5$\%$ to 1.5$\%$ as the pressure goes down. This decrease in the ESR contrast can again be attributed to an increased temperature. This is manifest in a significant shift in the zero-field splitting frequency as the pressure is reduced. As shown Fig. \ref{Pressure}-b), the temperature, deduced from Eq. 2, reaches 390 K at 0.2 mbars, consistent with the work of \cite{Toyli}. 
We note that the curve does not follow an inverse law dependence of temperature with pressure as in the experiment done in \cite{Hoang}.  Therer, a fit using
$T=T_0+\alpha/P$ was obtained, where $\alpha$ is a constant and $T_0$ the temperature at atmospheric pressure. This may be because in our work $\alpha$ depends upon the power and the pressure, most likely due to the damping-induced micromotion. 
A quantitative analysis would require the laser intensity seen by the NV$^-$ centers to be independent on pressure
or require measurements of the micro-motion amplitude, taking into account the achromatism of the aspherical lens and a correction of the position shift, which we will leave for further investigations.

To conclude, we observed electron spin resonances from NV centers with single diamond monocrystals levitating in a Paul trap under vacuum. Using the electronic spin resonance, we could observe heating at low vacuum pressures that can be attributed to the green laser beam used to read-out the electronic spin of the NV centers. 
Such NV thermometry method will be compared to the one used in \cite{Millen2} to derive the temperature of a levitating object from the center of mass temperature.
The heating is mostly due to the presence of nitrogen impurities in commercially available diamond powders \cite{Frangeskou} and could easily be overcome by employing purer diamonds made from Chemical Vapour Deposition or by using a pulsed laser scheme \cite{Neukirch}. With these improvements and reaching secular frequencies in the range of hundreds of kHz for the center of mass mode, this platform will be suited for hybrid-opto-mechanical schemes
 \cite{Scala, Yin, Yin2}. 

\begin{acknowledgements}
We would like to acknowledge fruitful discussions with the optics team at LPA and with L. Rondin.
This research has been partially funded by the French National Research Agency (ANR) through the project SMEQUI.
\end{acknowledgements}

\end{document}